\documentclass[prb,twocolumn,showpacs,superscriptaddress]{revtex4}%
\usepackage{amsfonts}
\usepackage{amsmath}
\usepackage{amssymb}
\usepackage{graphicx}%
\begin{document}

\title{Topological properties of Sb(111) surface: A density functional theory study}

%\author{xxx}
%\affiliation{x}
%\affiliation{xx}
%\affiliation{xxx}
%\affiliation{xxxx}
%\affiliation{xxxxx}
%\affiliation{xxxxxx}
%\affiliation{xxxxxxx}

\author{Shuang-Xi Wang}
\affiliation{State Key Laboratory for Superlattices and
Microstructures, Institute of Semiconductors, Chinese Academy of
Sciences, P. O. Box 912, Beijing 100083, People's Republic of China}
\affiliation{Department of Physics, Tsinghua University, Beijing
100084, People's Republic of China} \affiliation{LCP, Institute of
Applied Physics and Computational Mathematics, P.O. Box 8009,
Beijing 100088, People's Republic of China}
\author{Ping Zhang}
\thanks{Corresponding author; zhang\_ping@iapcm.ac.cn}
\affiliation{LCP, Institute of Applied Physics and Computational
Mathematics, P.O. Box 8009, Beijing 100088, People's Republic of
China}
\author{Shu-Shen Li}
\affiliation{State Key Laboratory for Superlattices and
Microstructures, Institute of Semiconductors, Chinese Academy of
Sciences, P. O. Box 912, Beijing 100083, People's Republic of China}

\pacs{68.55.Ln, 71.70.Ej, 73.20.At}

\date{\today}% It is always \today, today,

\begin{abstract}
By using first-principles plane wave calculations, we systematically
study the electronic properties of the thin film of antimony in
(111) orientation. Considering the spin-orbit interaction, for
stoichiometric surface, the topological states keep robust for six
bilayers, and can be recovered in the three bilayer film, which are
guarantied by time-reversal symmetry and inverse symmetry. For
reduced surface doped by Bi or Mn atom, localized 3-fold symmetric
features can be identified. Moreover, the non-trivial topological
states stand for non-magnetic substituted Bi atom, while can be
eliminated by adsorbed or substituted magnetic Mn atom.

\end{abstract}

\maketitle

\section{INTRODUCTION}

A new state of quantum matter-topological insulator (TI)-has
recently attracted great interest in condensed matter physics
\cite{Hasan2010,Qi2010}. The realization of TI HgTe both from
theoretical prediction and experimental observation
\cite{Bernevig2006,Konig2007}, open up opportunities for its
potential application in semiconductor spintronics. It is believed
that the quantum spin Hall effect as well as the time-reversal
symmetry plays an important role in the new material, protecting the
system from being disturbed by small perturbation caused by defects.
This urges people to desire for more other promising materials, such
as the Bi-based alloy Bi$_{1-x}$Sb$_x$, and layered compound
Bi$_{2}$X$_{3}$ (X=Se, Te)
\cite{Xia2009,Zhang2009,Chen2009,Tong2009}. Recently, the
semimetallic antimony has been drawing lots of attention by its
novel topological properties, and has become the proto-type system
of TI \cite{Sugawara2006,Gomes2009,Hsieh2009,Bian2011}.

Distinct from Bi, which is topologically trivial, according to the
parity of the band at the $\Gamma$ point \cite{Liu1995}, Sb has been
identified to be a strong TI \cite{Fu2007}. Experimentally, it has
been pointed out that the spin-split surface bands of Sb within its
bulk band gap are connected to the conduction band and valence band
\cite{Sugawara2006,Hsieh2009}. Moreover, the topologically
nontrivial Sb thin films exhibit novel properties and provide a
promising basis for spintronics applications such as device design
and integration \cite{Bian2011}.

Nevertheless, it is apparent that the detailed analysis of the
electronic structures of Sb is lacking, and there are still many
unanswered questions. Especially for the Sb(111) surface, it is much
desirable to identify the layer-dependence of the topological states
of Sb thin film, and the modulating of the surface properties by
impurities with or without magnetic moment \cite{Qi2008,Liu2009}.
Therefore, it is instructive to investigate the electronic
structures of Sb to deeply understand its topological properties.
This is not only of fundamental conceptual interest for
understanding the properties of TIs, but also paves the way for
realizing promising application of the new state of matter.

In this paper, we systematically study the properties of Sb(111)
surface by means of first-principles calculations. The bulk band
structure of Sb is presented to reveal the domination of spin-orbit
interaction (SOI) for topological properties of Sb. For the
stoichiometric Sb(111) surface, we investigate the layer-dependence
of the surface states of Sb thin film. Moreover, we calculate the
surface properties of Sb when impurities are introduced, including
substituted nonmagnetic Bi and magnetic Mn, as well as the adsorbed
Mn, and interesting results are obtained.

The rest of the paper is organized as follows. In Sec. II the
computational methods are briefly described. In Sec. III we present
and discuss our results for the bulk and surface properties of Sb.
Finally, in Sec. IV, we close our paper with conclusions of our main
results.

\section{COMPUTATIONAL METHODS}

The calculations are performed within density functional theory
using the Vienna \textit{ab-initio} simulation package (VASP)
\cite{VASP}. The PBE \cite{PBE} generalized gradient approximation
and the projector-augmented wave potential \cite{PAW} are employed
to describe the exchange-correlation energy and the electron-ion
interaction, respectively. Here the Sb 5\emph{s} and 5\emph{p}
electrons are treated as valence electrons. The SOI that has been
confirmed to play an important role in the electronic structure of
TI, is included during the calculation. The cutoff energy for the
plane wave expansion is set to 400 eV. And a Fermi broadening
\cite{Weinert1992} of 0.1 eV is chosen to smear the occupation of
the bands around the Fermi level by a finite-$T$ Fermi function and
extrapolating to $T$ = 0 K.

The Sb(111) surface is modeled by a slab composing of several(1-6)
bilayers (BL), and a vacuum region of 20 \AA. Obviously, existence
of the surface breaks the equivalence of the Sb atomic positions in
a layer, hence throughout this paper, we employ the notation Sb1 for
the first atomic plane, and Sb2 for the second atomic plane from the
surface, which are shown in Fig. 1(c). Integration over the
Brillouin zone is done using the Monkhorst-Pack scheme
\cite{Monkhorst1976} with 21$\times$21$\times$1 grid points for the
\emph{p}(1$\times$1) surface; for the \emph{p}(3$\times$3) surface,
7$\times$7$\times$1 grid points are used. The structures of slab are
fully optimized until the maximum residual ionic force is below 0.02
eV/\AA.

\section{RESULTS AND DISCUSSION}

\begin{figure}
\begin{center}
\includegraphics[width=0.8\linewidth]{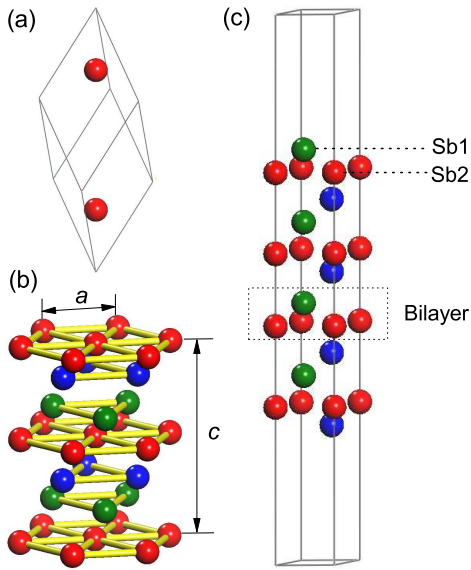}
\end{center}
\caption{(Color online) Atomic structure of bulk Sb and Sb(111)
surface. (a) Primitive cell of bulk Sb, (b) hexagonal unit cell of
bulk Sb, (c) slab model of Sb(111) surface. }
\label{fig1}
\end{figure}

The crystal structure of Sb is rhombohedral with the space group
$D_{3d}^{5}$($R\bar{3}m$), with two bismuth atoms in the trigonal
unit cell (see Fig. \ref{fig1}(a)). Equivalently, it can be
represented in terms of a hexagonally arranged layer structure,
which is shown in Fig. \ref{fig1}(b). The hexagonal unit cell has
three sets of bilayers, where each bilayer consists of two Sb atoms.
Structurally, bilayers in Sb form a stable unit with strong
intrabilayer bonds, while the interbilayer bonding is much weaker.
The calculated structure parameters are $a=4.392$ \AA, $c=11.432$
\AA, which are close to experimental data $a=4.3007$ \AA, $c=11.222$
\AA \cite{Barrett1963}. Moreover, the optimized internal parameter
is 0.233, consistent well with the experimental measurement 0.234.

\begin{figure}
\begin{center}
\includegraphics[width=0.8\linewidth]{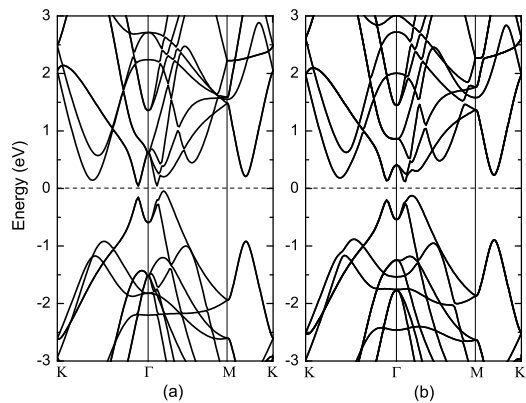}
\end{center}
\caption{Bulk band structure without SOI (a) and with SOI (b). The
Fermi level is set to zero.} \label{fig2}
\end{figure}

It has been proposed that Sb is a strong TI by calculating its Z$_2$
invariant $\nu$ (=1) from the knowledge of the parity of the
occupied Bloch wave function at the time-reversal invariant $\Gamma$
point in the Brillouin zone \cite{Fu2007}. This means that the SOI
dominates the electronic nature of Sb, which can be seen from the
band structure (see Fig. \ref{fig2}). It is clear that the band gap
is very small without SOI. While as the SOI is taken into
consideration, the band gap is enlarged up to about 0.237 eV.
Moreover, because of its topological electronic nature, we can
expect the existence of gapless surface states. Therefore, in the
following we will focus our attention on the properties of Sb(111)
surface.

\begin{figure}
\begin{center}
\includegraphics[width=0.8\linewidth]{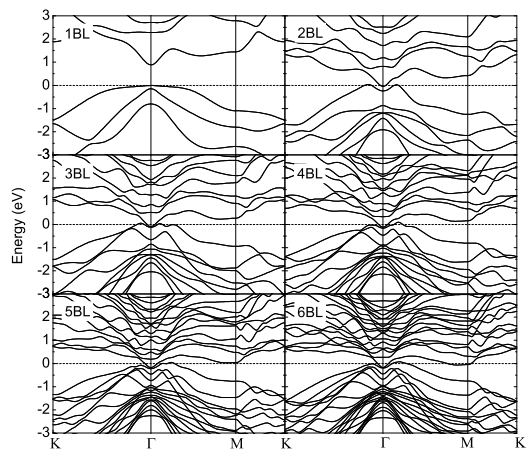}
\end{center}
\caption{Surface band structure with one to six bilayer thickness.
The Fermi level is set to zero.} \label{fig3}
\end{figure}

It is well known that the topological features are of strongly
thickness dependence of the TI thin film. For thinner film, the
coupling between top and bottom surfaces is strong enough to open up
a whole insulating gap. While with increasing thickness, the
inter-surface coupling becomes weaker and the topological features
will be recovered. This is the same case for Sb(111) surface. As
illustrated in Fig. \ref{fig3}, we show the evolution of band
structure of Sb films with the thickness from the single BL to six
BL. In a single BL film, the Sb electronic states (mainly from
5\emph{p} orbital) split into two parts forming a gap around the
Fermi level. The gap is as large as about 0.9 eV, implying the
strong coupling between two surfaces. In the case of two BL film,
the splitting of 5\emph{p} states decrease with the decline of the
coupling, leading to a semimetallic electronic structure. Obviously,
the topological features start to appear in the three BL case, where
a non-trivial helical edge state ($\nu_0$=1) below the Fermi level
at $\Gamma$ point can be identified. This helical state is consist
of two surface states degenerated at $\Gamma$ point but separated in
energy elsewhere by SOI \cite{Bian2011}. Nevertheless, it is
noticeable that for the case of four BL and five BL films, the gaps
are opened up again at $\Gamma$ point. This can be attributed to the
inverse asymmetry of the films with four and five BL. While the
topological states should be recovered when the number of BL of the
film is multiples of three, which conserves the inverse symmetry. As
expected, it can be seen that the topological states are recovered
for six BL. The double degenerate Sb(111) surface states contain a
single Dirac cone at the $\Gamma$ point, which is robust and
topologically protected by time-reversal symmetry. The Dirac point
is about 160 meV below the Fermi level, within the bulk band gap.
Compared with the experimental value of 230 meV \cite{Gomes2009},
the difference may arise from the subsurface defects observed in the
experiment.

Various impurities (magnetic or not) may have different impact on
the degeneracy and topological properties of Sb energy bands, which
will be illustrated in the following discussions based on the six BL
film. It has been shown that the alloy Bi$_{1-x}$Sb$_x$ is a 3D TI
\cite{Fu2007, Hsieh2008,Jeffrey2008}. Here it would be interesting
to investigate the surface state properties of reduced Sb(111)
surface with the existence of Bi as a nonmagnetic impurity.
Moreover, as a comparison, a magnetic impurity (Mn atom) is also
taken into consideration. During the calculations, the
\emph{p}(3$\times$3) surface is adopted with the substituted
impurities Bi or Mn atom taking the position of one subsurface Sb2
atom on each surface, while the adsorbed Mn atoms are symmetrically
located at the stable hcp sites above two surfaces of Sb(111).

\begin{figure}
\begin{center}
\includegraphics[width=0.8\linewidth]{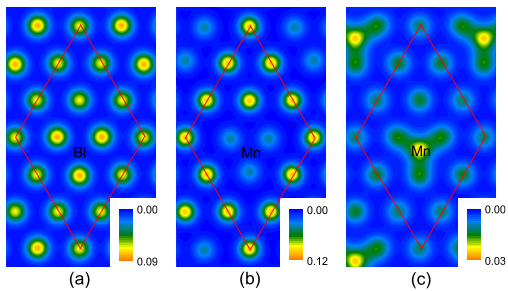}
\end{center}
\caption{(Color online) Surface charge density distributions at the
height of 2.0 \AA ~above the reduced Sb(111) surface: (a)
Bi-substituted, (b) Mn-substituted and (c) Mn-adsorbed.}
\label{fig4}
\end{figure}

Figure \ref{fig4} presents the surface charge density distributions
as simulation of STM topography at the height of 2.0 \AA ~above the
Bi-substituted, Mn-substituted and Mn-adsorbed surface,
respectively. One can see clearly that the localized 3-fold
symmetric features can be identified from all of three surfaces,
especially for the Mn-doped ones. For the Bi-substituted surface,
only tiny difference exists between the region above the Bi atom and
others, and this can be attributed to that Bi and Sb belong to the
same chemical group and possess the same number of valence
electrons. As a result the Bi-doped Sb(111) surface may maintain the
topological features, which will be illustrated later by its band
structure. For the Mn-doped surfaces, however, more evident
different features can be observed. It is noticeable that the charge
distribution is strongly depleted at the position just above the
substituted Mn atom, while the adsorbed Mn atom can be identified
clearly by the charge accumulation. Keeping in mind that the
magnetism of Mn atom, we will see that the magnetic impurity can
have significant effect on the surface states of Sb(111) surface.

\begin{figure}
\begin{center}
\includegraphics[width=0.8\linewidth]{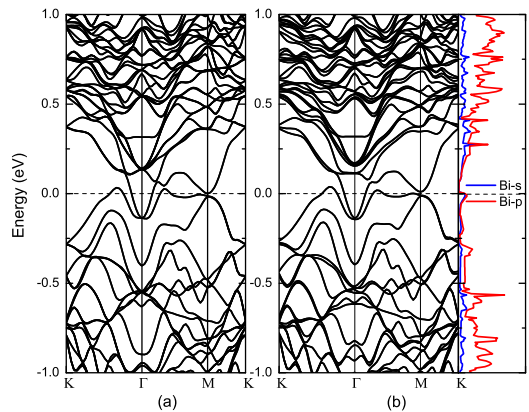}
\end{center}
\caption{(Color online) The band structure of clean (a) and Bi
substituted (b) Sb(111) surface.} \label{fig5}
\end{figure}

The band structure of Bi-doped Sb(111) surface is presented in Fig.
\ref{fig5}(b). For comparison, the band structure of stoichiometric
\emph{p}(3$\times$3) Sb(111) surface is also shown in Fig.
\ref{fig5}(a). It is clear that the topological surface state
remains to be robust despite of the existence of Bi impurity, and
the Dirac point of this reduced surface almost stays at the same
position as that of the stoichiometric surface. This can be
attributed to the following reasons. (i) The time-reversal symmetry
stands against nonmagnetic impurities. (ii) The energy bands of
Bi-doped system remain degenerate thus keep gapless. Moreover, from
the PDOS of Bi we can see that the doped Bi atom contributes little
to the band structure near the Fermi level and the Dirac point, thus
the topological features of Sb keep robust, which also consists with
the fact that the alloy Bi$_{1-x}$Sb$_x$ is a 3D TI inherited from
Sb \cite{Fu2007}.

\begin{figure}
\begin{center}
\includegraphics[width=0.8\linewidth]{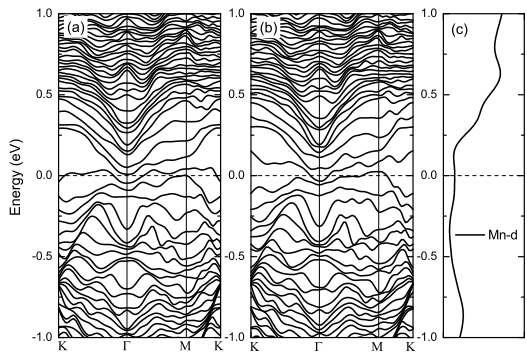}
\end{center}
\caption{The band structure of Mn-substituted Sb(111) surface: (a)
spin-up, (b) spin-down and (c) the PDOS of the Mn atomic $d$-bands.
} \label{fig6}
\end{figure}

\begin{figure}
\begin{center}
\includegraphics[width=0.8\linewidth]{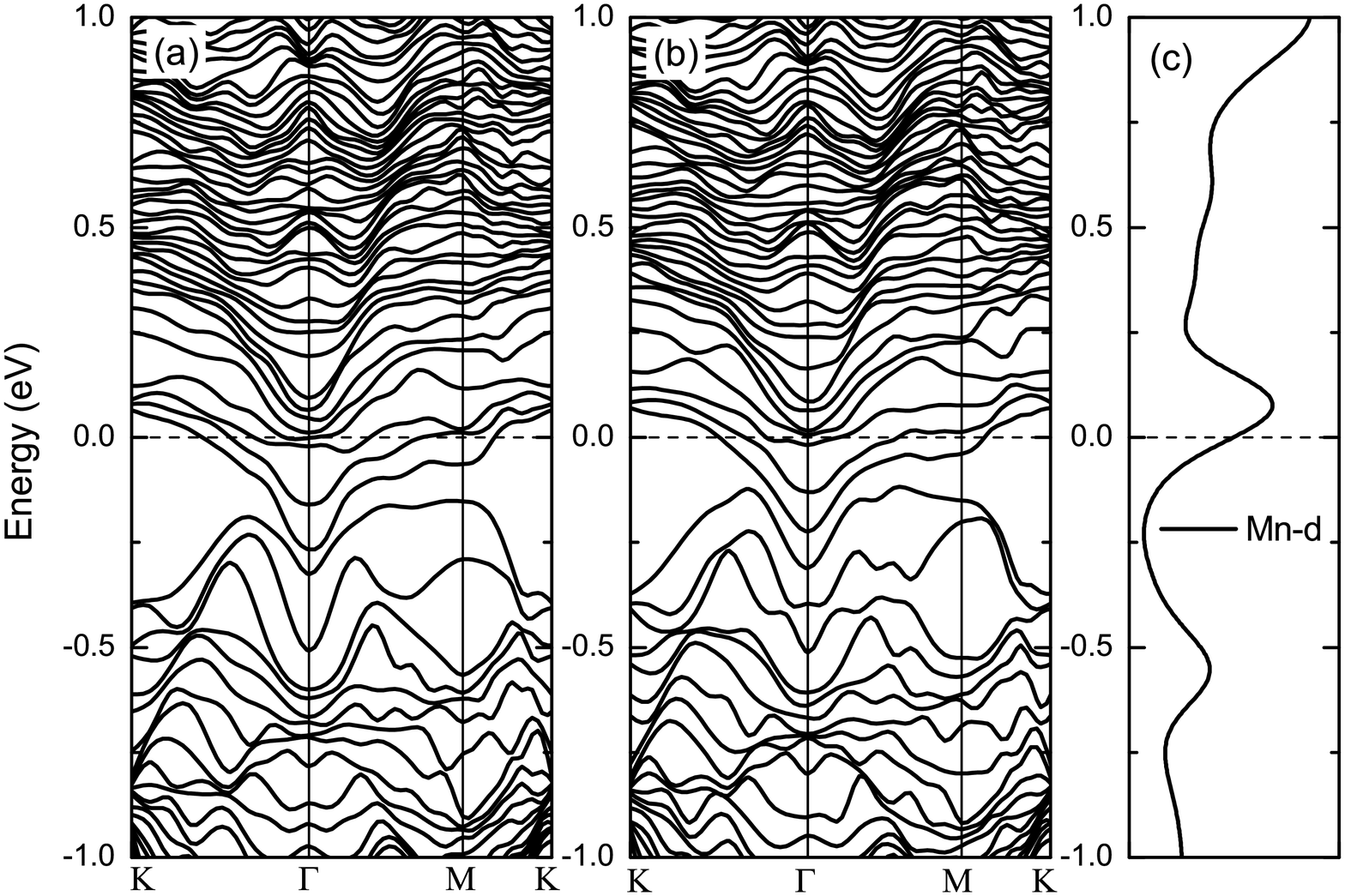}
\end{center}
\caption{The band structure of Mn-adsorbed Sb(111) surface: (a)
spin-up, (b) spin-down and (c) the PDOS of the Mn atomic $d$-bands.
} \label{fig7}
\end{figure}

Theoretically, if the impurity carries a magnetic moment, the
time-reversal symmetry is explicitly broken, and a local energy gap
will be opened up at the Dirac point \cite{Liu2009}. While up to now
relevant experimental observations are lack yet, therefore it is
desirable to have a direct and intuitional sight into it. By
calculations we find that the magnetic impurities can obviously lift
the degeneracy protected by the time-reversal symmetry. For clarity,
here we plot the lifted band structures of Mn-substituted and
adsorbed Sb(111) surfaces dividedly in (a) and (b) of Fig.
\ref{fig6} and Fig. \ref{fig7}, labeled by spin-up and spin-down,
respectively. It can be seen that both of the magnetic impurities
eliminate the Dirac point by opening up a gap at $\Gamma$ point.
From the PDOS (See Fig. \ref{fig6}(c) and Fig. \ref{fig7}(c)), we
find that unlike the doped Bi atom, the magnetic \emph{d} orbital of
Mn atom contributes much to the states near the Dirac point, hence
clearly breaks the time-reversal symmetry. As expected
\cite{Liu2009} a ferromagnetic ground state is formed on the TI
surface by the introduced magnetic Mn impurity.

\section{CONCLUSIONS}

In conclusion, we have systematically studied the topological
properties of Sb(111) surface by density functional calculation. We
found that the stoichiometric Sb(111) surface possesses single Dirac
point protected by the time-reversal symmetry and inverse symmetry.
And the topological states are layer dependent and keep robust for
six bilayers film. Moreover, we revealed that the non-trivial
topological states stand for non-magnetic substituted Bi, while the
substituted or adsorbed magnetic Mn atom can obviously eliminate the
Dirac point. We expect the present results for the topological
features of Sb are greatly helpful for understanding and application
of the topological insulator.

\end{document}